\documentclass{PoS}

\title{Measuring polarizations of bottom, charm, strange, up and down quarks in top decays}

\ShortTitle{Measuring polarizations of bottom, charm, strange, up and down quarks in top decays}

\author{Yevgeny Kats\\
        Department of Particle Physics and Astrophysics\\
        Weizmann Institute of Science\\
        Rehovot 7610001, Israel\\
        E-mail: \email{yevgeny.kats@weizmann.ac.il}}

\abstract{Standard Model $t\bar t$ samples in ATLAS and CMS provide an opportunity to conduct measurements of the polarization of quarks produced in top-quark decays. The bottom quarks from the $t\to W^+b$ decays, as well as the charm, strange, up and down quarks from the subsequent $W^+ \to c\bar s, u\bar d$ decays, are polarized. When a polarized quark hadronizes to a baryon, its polarization is partly preserved, and is reflected in the angular distributions of the baryon decay products. For the strange and bottom quarks, this effect has already been seen in hadronic $Z$ decays at LEP. In Run~2 of the LHC, $t\bar t$ samples will have comparable statistics and several advantages from the point of view of flavor tagging. We propose strategies for such measurements in ATLAS and CMS. We estimate that precision of order 10\% is attainable for the polarizations of the strange, charm and bottom quarks. In future higher-statistics runs, polarizations of up and down quarks might be measurable as well.}
 
\FullConference{8th International Workshop on Top Quark Physics\\
                  14-18 September, 2015\\
                  Ischia, Italy}

\begin{document}

An ability to measure the spin state (polarization) of a quark produced in a hard process at the LHC can turn out useful for characterizing, or perhaps even discovering, new physics. Indeed, information about new physics is typically encoded in the Standard Model particles produced in the event. Since a quark carries information in both its momentum and polarization, it is desirable to be able to measure both. Quark momentum is easily reconstructed in ATLAS and CMS by measuring the jet. Is it possible to also reconstruct quark polarization? We have argued~\cite{Galanti:2015pqa,Kats:2015cna} that the answer is positive. This brief note summarizes our proposals for measurements of $s$, $c$ and $b$ quark polarizations in the ATLAS and CMS $t\bar t$ samples of Run~2. These analyses will calibrate and validate the measurement techniques, thus preparing them for new-physics applications. Additionally, along with analogous measurements of $u$ and $d$ quark polarizations in the longer run, they will provide information on interesting QCD quantities.

Measurements of polarization of top quarks themselves are by now quite standard. Since the top decays before hadronization, its polarization can be extracted straightforwardly from its decay angular distributions. As expected, tops are found to be polarized in the electroweak single-top production and unpolarized in the QCD-dominated $t\bar t$ production (see, e.g.,~\cite{CMS-top-polarization}). If a new-physics source of top quarks is discovered, measuring their polarization will analogously teach us about their production mechanism. We asked whether this could be done also with other quarks: $c$ and $b$ in~\cite{Galanti:2015pqa} and $u$, $d$ and $s$ in~\cite{Kats:2015cna}.

Since quarks are observed as jets of hadrons, extracting the quark polarization is not straightforward. Due to the complicated nature of QCD, it is a priori unclear whether the polarization survives the hadronization process and the subsequent decays and which hadron(s) carry it. However, for heavy quarks, $m_q \gg \Lambda_{\rm QCD}$, like the $b$ and $c$, two simplifications arise. First, the jet usually contains a very energetic heavy-flavored hadron that carries the original quark (see, e.g.,~\cite{FFs}). Second, when this hadron is a baryon, it is expected to retain a fixed ${\cal O}(1)$ fraction of the quark polarization~\cite{Falk:1993rf} (see also~\cite{Galanti:2015pqa}). This polarization can be extracted from kinematic distributions of the baryon decay products. Evidence of this effect has been observed in $\Lambda_b$ ($\approx bud$) baryons in $Z\to b\bar b$ samples at LEP~\cite{LEP-Lambda_b}. Moreover, sizable polarization was observed also for $\Lambda$ ($\approx sud$) baryons in $Z \to q\bar q$, for $\Lambda$'s carrying a large fraction, $z \gtrsim 0.3$, of the quark momentum~\cite{LEP-Lambda}. This implies ${\cal O}(1)$ $s$-quark polarization retention in energetic $\Lambda$'s (see~\cite{Kats:2015cna} and references therein).

At the LHC, a great source of polarized quarks is $t\bar t$ samples. The decay $t\to W^+b$ produces polarized $b$ quarks, and the subsequent $W^+\to c\bar s$, $u\bar d$ decays produce polarized $c$, $s$, $u$, $d$ quarks. It is easy to select a clean $t\bar t$ sample (e.g., in the lepton + jets channel), and jets of different quark flavors can be studied separately using kinematic event reconstruction and $b$ and $c$ tagging. From the point of view of statistics, already in Run~2 of the LHC, the $t\bar t$ events will provide as large samples of polarized quarks as the $Z$ decays at LEP. 

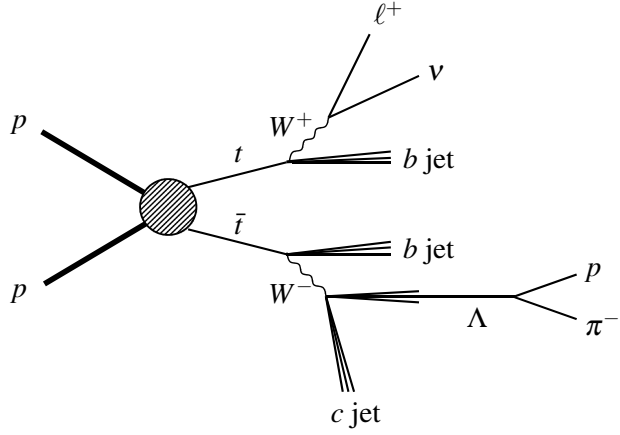
\begin{figure}[t]
\begin{center}
\vspace{-3mm}
\includegraphics[scale=0.35]{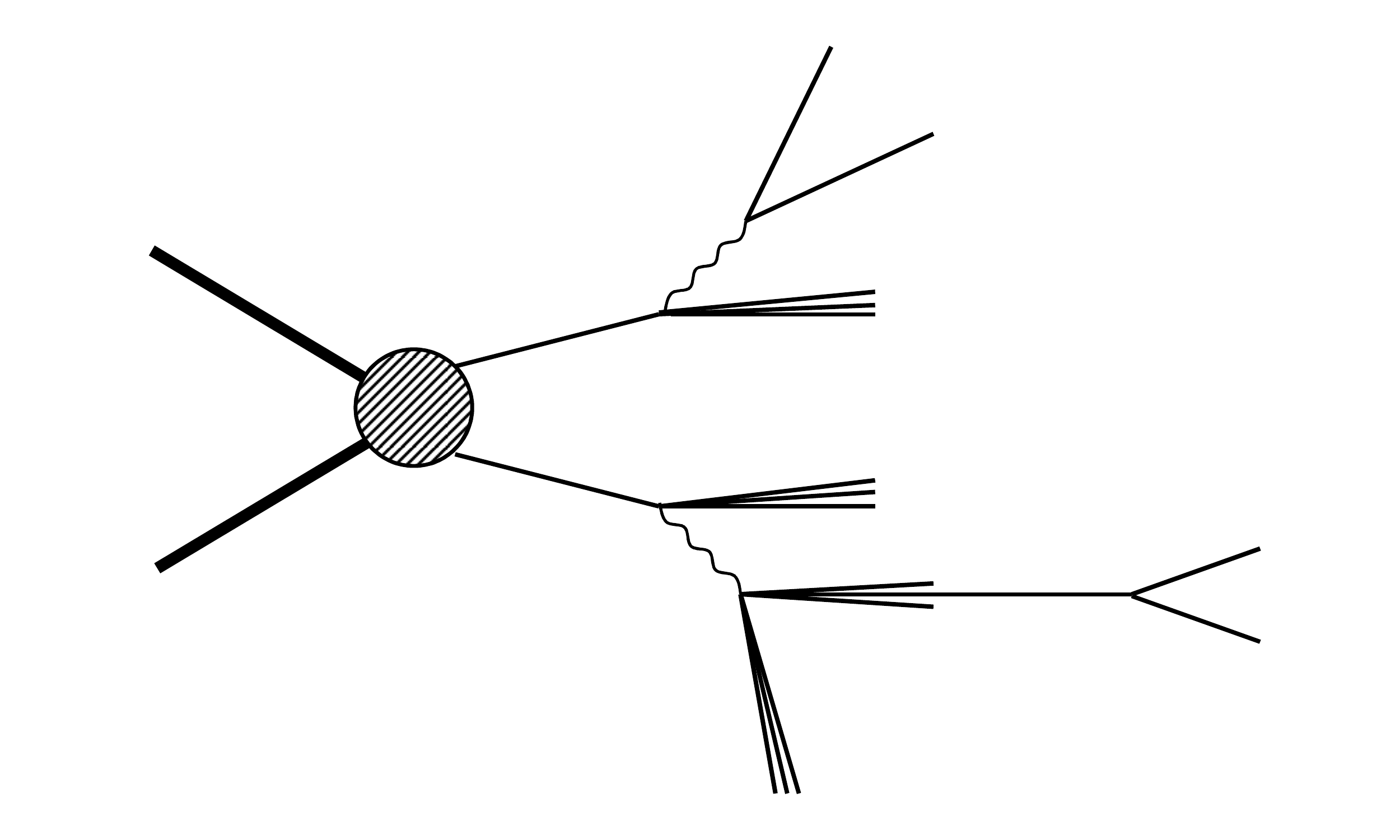}
\begin{picture}(0,0)(0,0)
\put(-240,108){$p$}
\put(-240,44){$p$}
\put(-156,95){$t$}
\put(-156,68){$\bar t$}
\put(-143,42){$W^-$}
\put(-143,105){$W^+$}
\put(-104,147){$\ell^+$}
\put(-84,126){$\nu$}
\put(-93,93){$b$ jet}
\put(-93,59){$b$ jet}
\put(-120,-3){$c$ jet}
\put(-69,33){$\Lambda$}
\put(-25,52){$p$}
\put(-25,30){$\pi^-$}
\end{picture}
\caption{Measuring polarization of $s$ quarks from $W^- \to \bar c s$ decays using $\Lambda$ decays in $t\bar t$ samples. The $s$ jets are identified by tagging the accompanying $c$ jets using generic charm tagging methods.}
\label{fig:ttbar-chain-Lambda}
\end{center}
\end{figure}

The fragmentation fractions to the baryons of interest are $f(s\to\Lambda) \approx 3\%$ for $z > 0.3$~\cite{Albino:2008fy}, $f(c\to\Lambda_c^+) \approx 6\%$~\cite{Lisovyi:2015uqa}, $f(b\to\Lambda_b) \approx 7\%$~\cite{Galanti:2015pqa}. While in principle many of the baryon decays can be used for the polarization measurements, we have focused on $\Lambda \to p\pi^-$ ($\mbox{BR} \approx 64\%$, $\alpha \approx 0.64$), $\Lambda_c^+ \to p K^-\pi^+$ ($\mbox{BR} \approx 7\%$, $\alpha_{K^-} \sim {\cal O}(1)$) and $\Lambda_b \to X_c\mu^-\bar\nu_\mu$ ($\mbox{BR} \approx 10\%$, $\alpha_{\bar\nu_\mu} \approx 1$),\footnote{$X_c$ stands for any collection of particles containing a charmed hadron (typically a $\Lambda_c^+$, but sometimes a $D$ meson). We considered a fully inclusive selection of $X_c$, which admits a large intrinsic background from semileptonic $B$-meson decays, as well as a semi-inclusive selection that requires the presence of a reconstructed $\Lambda \to p\pi^-$ decay, and an exclusive selection that focuses on several fully reconstructible $\Lambda_c^+$ decay modes. Sensitivities turned out comparable.} where the numbers in parentheses give the branching ratio and the spin analyzing power of each decay. 

\begin{figure}[b]
\begin{center}
\vspace{-3mm}
\includegraphics[scale=0.35]{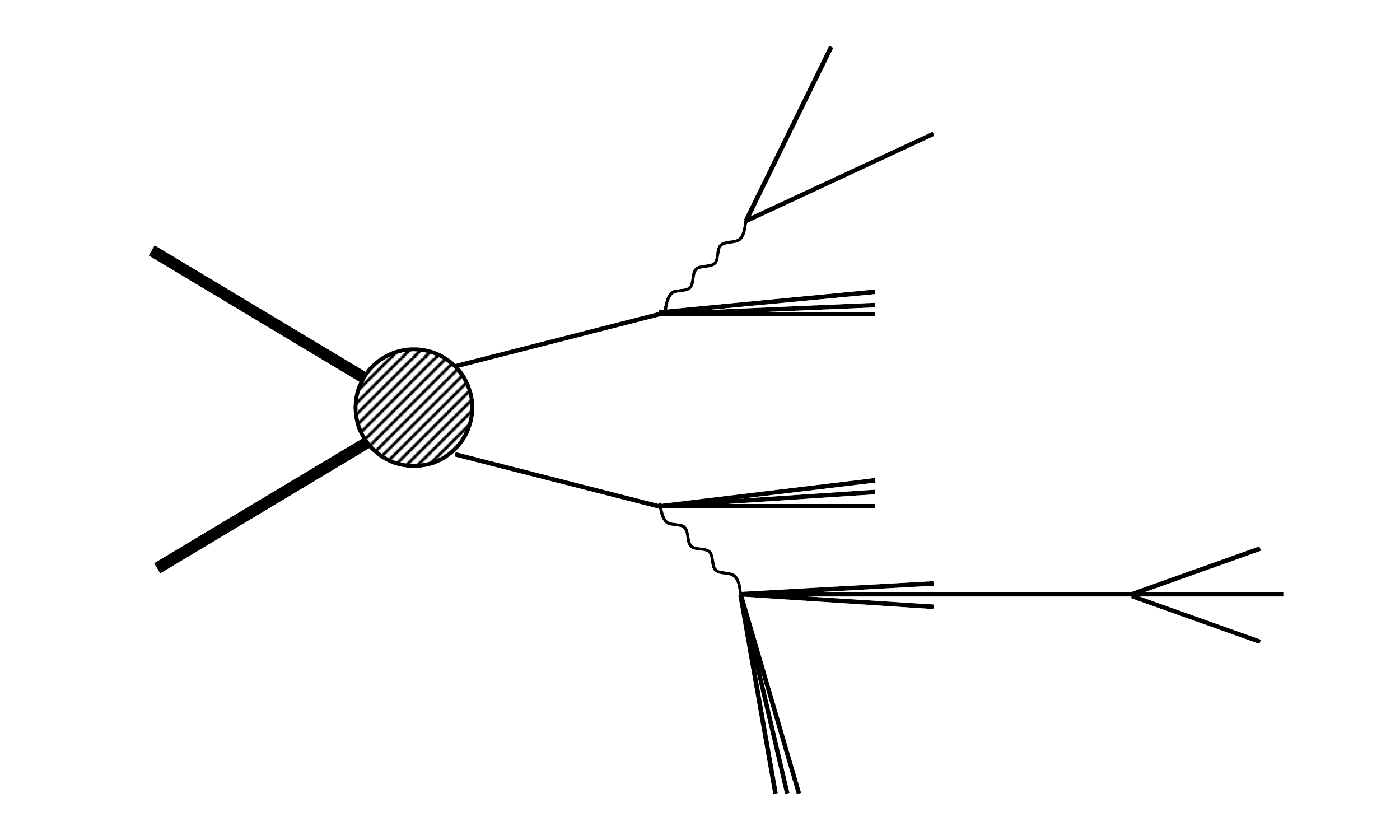}
\begin{picture}(0,0)(0,0)
\put(-240,108){$p$}
\put(-240,44){$p$}
\put(-156,94){$\bar t$}
\put(-156,68){$t$}
\put(-143,42){$W^+$}
\put(-143,103){$W^-$}
\put(-104,147){$\ell^-$}
\put(-84,126){$\bar\nu$}
\put(-93,93){$b$ jet}
\put(-93,59){$b$ jet}
\put(-118,-3){jet}
\put(-75,33){$\Lambda_c^+$}
\put(-25,54){$p$}
\put(-20,41){$K^-$}
\put(-25,29){$\pi^+$}
\end{picture}
\caption{Measuring polarization of $c$ quarks from $W^+ \to c \bar s$ decays using $\Lambda_c^+$ decays in $t\bar t$ samples.}
\label{fig:ttbar-chain-Lambda_c}
\end{center}
\end{figure}
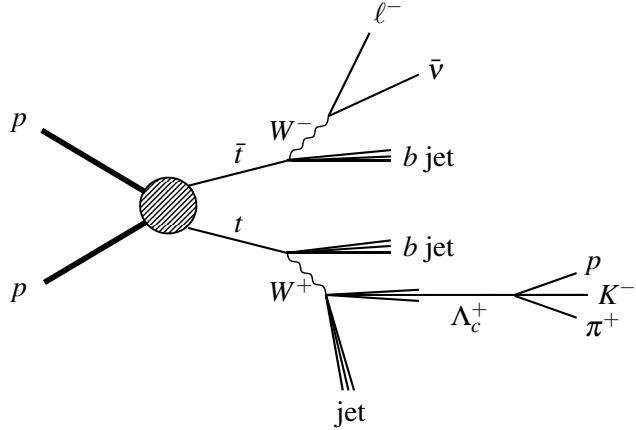

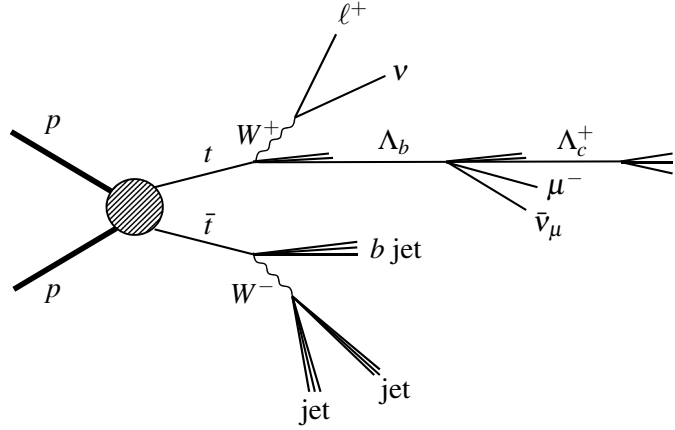
\begin{figure}[t]
\begin{center}
\vspace{-3mm}
\includegraphics[scale=0.35]{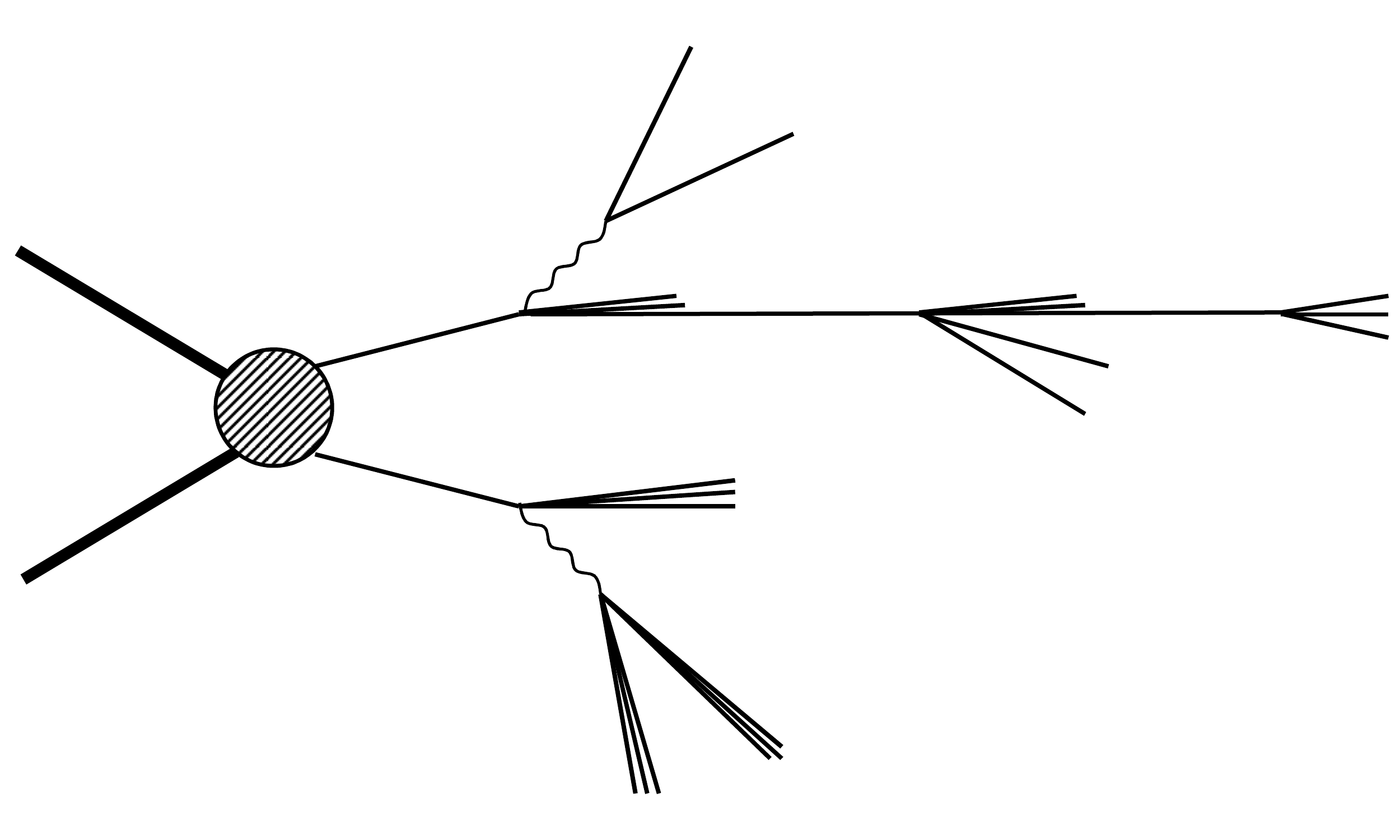}
\begin{picture}(0,0)(0,0)
\put(-240,108){$p$}
\put(-240,44){$p$}
\put(-180,95){$t$}
\put(-180,68){$\bar t$}
\put(-170,42){$W^-$}
\put(-168,102){$W^+$}
\put(-130,147){$\ell^+$}
\put(-110,126){$\nu$}
\put(-118,59){$b$ jet}
\put(-144,-2){jet}
\put(-114,6){jet}
\put(-115,100){$\Lambda_b$}
\put(-48,100){$\Lambda_c^+$}
\put(-52,83){$\mu^-$}
\put(-57,69){$\bar\nu_\mu$}
\end{picture}
\caption{Measuring polarization of $b$ quarks from $t \to W^+ b$ decays using $\Lambda_b$ decays in $t\bar t$ samples.}
\label{fig:ttbar-chain-Lambda_b}
\end{center}
\end{figure}

All the measurements start with a standard $t\bar t$ selection and kinematic reconstruction. The baryon decays are then selected and reconstructed in the jets of interest, as illustrated in figures~\ref{fig:ttbar-chain-Lambda}--\ref{fig:ttbar-chain-Lambda_b}. The polarization is extracted from the angular distributions of the baryon decay products in the baryon rest frame.

We find (see~\cite{Galanti:2015pqa,Kats:2015cna} for many additional details) that these methods will allow measuring the polarizations of the $s$, $c$ and $b$ quarks in Run~2 with ${\cal O}(10\%)$ precisions. In practice, these measurements will determine the ratios of the baryon and the (known) initial quark polarizations. These quantities will be useful for interpreting quark polarization measurements in new-physics samples (see example in~\cite{Kats:2015cna}) and for testing QCD models.

While $u$ and $d$ quark polarizations obviously cannot be measured using decays of protons or neutrons, one may again consider the $\Lambda$ ($\approx sud$), as in the $s$-quark case. Even though in the na\"{i}ve quark model the spin of the $\Lambda$ is carried entirely by the $s$ quark, the SU(3) flavor symmetry applied to the nucleon DIS data suggests that the $u$ and $d$ quarks each carry about $-20\%$ of the $\Lambda$ spin~\cite{Burkardt-Jaffe}.\footnote{Further inputs may come from polarized DIS and polarized $pp$ collisions (e.g.,~\cite{polarized-exp}) and lattice QCD (e.g.,~\cite{LQCD}).} Still, the correspondingly suppressed polarization transfer, as well as the smallness of $f(u\to\Lambda)$ and $f(d\to\Lambda)$ relative to $f(s\to\Lambda)$ (e.g., by factors of $\sim 4$ for $z > 0.3$~\cite{Albino:2008fy}) and the expected large contamination from $W^+\to c\bar s$ decays (as a $c$-tag veto is only partially effective), imply that statistics beyond Run~2 will be required for the $u$ and $d$ quark polarization measurements in $t\bar t$ samples.

\end{document}